# Generic Expression Hardness Results for Primitive Positive Formula Comparison

Simone Bova[1], Hubie Chen[2], and Matthew Valeriote[3]

[1] Department of Mathematics
Vanderbilt University (Nashville, USA)
simone.bova@vanderbilt.edu
[2] Departament de Tecnologies de la Informació i les Comunicacions
Universitat Pompeu Fabra (Barcelona, Spain)
hubie.chen@upf.edu
[3] Department of Mathematics & Statistics
McMaster University (Hamilton, Canada)
matt@math.mcmaster.ca

**Abstract.** We study the expression complexity of two basic problems involving the comparison of primitive positive formulas: equivalence and containment. In particular, we study the complexity of these problems relative to finite relational structures. We present two generic hardness results for the studied problems, and discuss evidence that they are optimal and yield, for each of the problems, a complexity trichotomy.

## 1 Introduction

**Overview.** A *primitive positive (pp)* formula is a first-order formula defined from atomic formulas and equality of variables using conjunction and existential quantification. The class of primitive positive formulas includes, and is essentially equivalent to, the class of *conjunctive queries*, which is well-established in relational database theory as a pertinent and useful class of queries, and which has been studied complexity-theoretically from a number of perspectives (see for example [19,18,1]). In this paper, we study the complexity of the following fundamental problems, each of which involves the comparison of two pp-formulas $\phi, \phi'$ having the same free variables, over a relational structure.

- Equivalence: are the formulas $\phi, \phi'$ equivalent–that is, do they have the same satisfying assignments–over the structure?
- Containment: are the satisfying assignments of $\phi$ contained in those of $\phi'$, over the structure?

We study the complexity of these computational problems with respect to various fixed structures. That is, we parameterize each of these problems with respect to the structure to obtain a family of problems, containing one member for each structure, and study the resulting families of problems. To employ the terminology of Vardi [20], we study the *expression complexity* of the presented comparison tasks. The suggestion here is that various relational structures–which may represent databases or knowledge bases, according to use–may possess structural characteristics that affect the complexity of the resulting problems, and our interest is in understanding this interplay. The present work focuses on relational structures that are finite (that is, have finite universe), and we assume that the structures under discussion are finite.

In this paper, we present two general expression hardness results on the problems of interest. In particular, each of our two main results provides a sufficient condition on a structure so that the problems are hard for certain complexity classes. Furthermore, we give evidence that our results are optimal, in that the conditions that they involve in fact describe dichotomies in the complexity of the studied problems; put together, our results indicate, for each of the studied problems, a complexity trichotomy.

Our study utilizes universal-algebraic tools that aid in understanding the set of primitive positively definable relations over a given structure. It is known that, relative to a structure, the set of relations that are definable by a primitive positive formula forms a robust algebraic object known as a *relational clone*; a known Galois correspondence associates, in a bijective manner, each such relational clone with a *clone*, a set of operations with certain closure properties. This correspondence provides a way to pass from a relational structure **B** to an algebra $\mathbb{A}_\mathbf{B}$ whose set of operations is the mentioned clone, in such a way that two structures having the same algebra have the same complexity (for each of the mentioned problems). In a previous paper by the present authors [6], we developed this correspondence and presented some basic complexity results for the problems at hand, including a classification of the complexity of the problems on all two-element structures.

**Our hardness results.** Our first hardness result (Section 3) yields that for any structure **B** whose associated algebra $\mathbb{A}_\mathbf{B}$ gives rise to a variety $\mathcal{V}(\mathbb{A}_\mathbf{B})$ that *admits the unary type*, both the equivalence and containment problems are $\Pi_2^p$-hard. Note that this is the maximal complexity possible for these problems, as the problems are contained in the class $\Pi_2^p$. The condition of admitting the unary type originates from tame congruence theory, a theory developed to understand the structure of finite algebras [13]. We observe that this result implies a dichotomy in the complexity of the studied problems under the *G-set conjecture* for the constraint satisfaction problem (CSP), a conjecture put forth by Bulatov, Jeavons, and Krokhin [7] which predicts exactly where the tractability/intractability dichotomy lies for the CSP. (Recall that the CSP can be formulated as the problem of deciding, given a structure **B** and a primitive positive sentence $\phi$, whether or not $\phi$ holds on **B**.) In particular, under the G-set conjecture, the structures not obeying the described condition have equivalence and containment problems in coNP. The resolution of the G-set conjecture, on which there has been focused and steady progress over the past decade [9,14,10,3], would thus, in combination with our hardness result, yield a coNP/$\Pi_2^p$-complete dichotomy for the equivalence and containment problems. In fact, our hardness result already unconditionally implies dichotomies for our problems for all classes of structures where the G-set conjecture has already been established, including the class of three-element structures [9], and the class of conservative structures [8].

One formulation of the G-set conjecture is that, for a structure **B** whose associated algebra $\mathbb{A}_\mathbf{B}$ is idempotent, the absence of the unary type in the variety generated by $\mathbb{A}_\mathbf{B}$ implies that the CSP over **B** is polynomial-time tractable. The presence of the unary type is a known sufficient condition for intractability in the idempotent case [7,10], and this conjecture predicts exactly where the tractability/intractability dichotomy lies for the CSP. It should be noted, however, that the boundary that is suggested by our hardness result for the equivalence and containment problems is *not* the same as the boundary suggested by the G-set conjecture for the CSP. The G-set conjecture, which is typically phrased on idempotent algebras, yields a prediction on the CSP complexity of all structures via a theorem [7] showing that each structure **B** has the same CSP complexity as a structure **B**$'$ whose associated algebra is idempotent. The mapping from **B** to **B**$'$ does not preserve the complexity of the problems studied here, and indeed, there are examples of two-element structures **B** such that our hardness result applies to **B**–the equivalence and containment problems on **B** are $\Pi_2^p$-complete–but **B**$'$ does not admit the unary type and indeed has a polynomial-time tractable CSP [6]. Our new result requires establishing a deeper understanding of the identified algebras' structure, some of which admit a tractable CSP, in order to obtain hardness.

Our second hardness result (Section 4) shows that for any structure **B**, if the variety $\mathcal{V}(\mathbb{A}_\mathbf{B})$ is not congruence modular, then the equivalence and containment problems are coNP-hard. Previous work identified one most general condition for the tractability of the equivalence and containment problems: if the algebra *has few subpowers*–a combinatorial condition [4,14] involving the number of subalgebras of powers of an algebra–then these problems are polynomial-time tractable [6, Theorem 7]. This second hardness result appears to perfectly complement this tractability result: there are no known examples of algebras $\mathbb{A}_\mathbf{B}$ (of structures **B** having finitely many relations) that are not covered by one of these results, and in fact the *Edinburgh conjecture* predicts that none exist, stating that every such algebra $\mathbb{A}_\mathbf{B}$ that generates a congruence modular variety also has few subpowers. Concerning this conjecture, it should be pointed out that the resolution of the *Zadóri*

*conjecture*, a closely related conjecture of which the Edinburgh conjecture is a generalization, was recently announced by Libor Barto [2]. The Edinburgh conjecture is of current interest, with recent work presented by Ralph McKenzie and colleagues. We also point out that this conjecture (as with the Zadóri conjecture) is purely algebraic, making no references to notions of computation.

In summary, up to polynomial-time computation, **we completely resolve the complexity of the studied problems on all finite structures, showing a P/coNP-complete/$\Pi_2^p$-complete trichotomy–modulo two conjectures;** one is computational and one is algebraic, and for each there is both highly non-trivial supporting evidence and current investigation. The coNP/$\Pi_2^p$-complete dichotomy is presented in Section 3 (see Theorems 5 and 6) and the P/coNP-hard dichotomy is presented in Section 4 (see Theorems 11 and 12).

## 2 Preliminaries

Here, a *signature* is a set of relation symbols, each having an associated arity; we assume that all signatures are of finite size. A *relational structure* over a signature $\sigma$ consists of a universe $B$ and, for each relation symbol $R \in \sigma$, a relation $R^{\mathbf{B}} \subseteq B^k$ where $k$ is the arity of $R$. We assume that all relational structures under discussion have universes of finite size. A *primitive positive formula (pp-formula)* on $\sigma$ is a first-order formula formed using equalities on variables ($x = x'$), atomic formulas $R(x_1, \ldots, x_k)$ over $\sigma$, conjunction ($\wedge$), and existential quantification ($\exists$).

We now define the problems that will be studied.

**Definition 1.** *We define the following computational problems; in each, an instance consists of a relational structure $\mathbf{B}$ and a pair $(\phi, \phi')$ of pp-formulas over the signature of $\mathbf{B}$ having the same set of free variables $X$.*

- PPEQ: *decide if $\phi$ and $\phi'$ are equivalent, that is, whether for all $f : X \to B$, it holds that $\mathbf{B}, f \models \phi$ iff $\mathbf{B}, f \models \phi'$.*
- PPCON: *decide if $\phi$ is contained in $\phi'$, that is, whether for all $f : X \to B$, it holds that $\mathbf{B}, f \models \phi$ implies $\mathbf{B}, f \models \phi'$.*

*For every relational structure $\mathbf{B}$, we define PPEQ($\mathbf{B}$) to be the problem PPEQ where the structure is fixed to be $\mathbf{B}$; hence, an instance of PPEQ($\mathbf{B}$) is just a pair $(\phi, \phi')$ of pp-formulas. We define the family of problems PPCON($\mathbf{B}$) analogously.*

We now identify some basic complexity properties of these problems. First, the PPEQ and PPCON problems are contained in $\Pi_2^p$; this is straightforward to verify. Next, there is a direct reduction from PPCON($\mathbf{B}$) to PPEQ($\mathbf{B}$). Throughout the paper, the notion of reduction used is polynomial-time many-one reducibility.

**Proposition 1** *For each structure $\mathbf{B}$, the problem PPCON($\mathbf{B}$) reduces to the problem PPEQ($\mathbf{B}$).*

**Proof.** The reduction, given an instance $(\phi, \phi')$ of PPCON($\mathbf{B}$), outputs the instance $(\phi, \phi \wedge \phi')$ of PPEQ($\mathbf{B}$). □

We now review the relevant algebraic concepts to be used. An *algebra* is a pair $\mathbb{A} = (A, F)$ such that $A$ is a nonempty set, called the *domain* or *universe* of the algebra, and $F$ is a set of finitary operations on $A$. Let $\mathbb{A} = (A, F)$ be an algebra; a *term operation* of $\mathbb{A}$ is a finitary operation obtained by composition of (1) operations in $F$ and (2) projections on $A$, and a *polynomial operation* is a finitary operation obtained by composition of (1) operations in $F$, (2) projections on $A$ and (3) constants from $A$. An operation $f(x_1, \ldots, x_n)$ on A is said to be *idempotent* if the equality $f(a, a, \ldots, a) = a$ holds for all $a \in A$. An algebra $\mathbb{A}$ is *idempotent* if all of its term operations are.

Let $B$ be a nonempty set, let $f$ be an $n$-ary operation on $B$, and let $R$ be a $k$-ary relation on $B$. We say that $f$ *preserves* $R$ (or $f$ is a *polymorphism* of $R$), if for every length $n$ sequence of tuples $t_1, \ldots, t_n \in R$, denoting the tuple $t_i$ by $(t_{i,1}, \ldots, t_{i,k})$, it holds that the tuple

$$f(t_1, \ldots, t_n) = (f(t_{1,1}, \ldots, t_{n,1}), \ldots, f(t_{1,k}, \ldots, t_{n,k}))$$

is in $R$. We say that a relation $R$ is *compatible* with a set of operations if it is preserved by all of the operations. We extend this terminology to relational structures: an operation $f$ is a polymorphism of a relational structure $\mathbf{B}$ if $f$ is a polymorphism of every relation of $\mathbf{B}$. We use $\mathsf{Pol}(\mathbf{B})$ to denote the set of all polymorphisms of a relational structure $\mathbf{B}$, and use $\mathbb{A}_\mathbf{B}$ to denote the algebra $(B, \mathsf{Pol}(\mathbf{B}))$. Dually, for an operation $f$, we use $\mathsf{Inv}(f)$ to denote the set of all relations that are preserved by $f$, and for a set of operations $F$, we define $\mathsf{Inv}(F)$ as $\bigcap_{f \in F} \mathsf{Inv}(f)$. We will make use of the following result connecting the $\mathsf{Pol}(\cdot)$ and $\mathsf{Inv}(\cdot)$ operators to pp-definability.

**Theorem 2.** *(Geiger [12]/Bodcharnuk et al. [5]) Let $\mathbf{B}$ be a finite relational structure. The set of relations $\mathsf{Inv}(\mathsf{Pol}(\mathbf{B}))$ is equal to the set of relations that are pp-definable over $\mathbf{B}$.*

We associate to each algebra $\mathbb{A} = (A, F)$ a set of problems $\mathsf{PPEQ}(\mathbb{A})$, namely, the set containing all problems $\mathsf{PPEQ}(\mathbf{B})$ where $\mathbf{B}$ has universe $A$ and $F \subseteq \mathsf{Pol}(\mathbf{B})$. We define $\mathsf{PPCON}(\mathbb{A})$ similarly. For a complexity class $\mathcal{C}$, we say that the problem $\mathsf{PPEQ}(\mathbb{A})$ is $\mathcal{C}$-hard if $\mathsf{PPEQ}(\mathbb{A})$ contains a problem $\mathsf{PPEQ}(\mathbf{B})$ that is $\mathcal{C}$-hard. We define $\mathcal{C}$-hardness similarly for $\mathsf{PPCON}(\mathbb{A})$.

**Theorem 3.** *Let $\mathbf{B}$ be a finite relational structure, and let $\mathcal{C}$ be a complexity class closed under polynomial-time many-one reductions. The problem $\mathsf{PPEQ}(\mathbf{B})$ is $\mathcal{C}$-hard if and only if $\mathsf{PPEQ}(\mathbb{A}_\mathbf{B})$ is $\mathcal{C}$-hard. The same result holds for $\mathsf{PPCON}(\cdot)$.*

**Proof**. The proof of [6, Theorem 2] applies to each of the problems. □

The notion of a *variety* is typically defined on indexed algebras; a variety is a class of similar algebras that is closed under the formation of homomorphic images, subalgebras, and products. For our purposes here, however, we may note that the variety generated by an algebra $\mathbb{A}$, denoted by $\mathcal{V}(\mathbb{A})$, is known to be equal to $HSP(\{\mathbb{A}\})$, where the operator $H$ (for instance) is the set of algebras derivable by taking homomorphic images of algebras in the given argument set.

**Theorem 4.** *Suppose that $\mathbb{B} \in \mathcal{V}(\mathbb{A})$. Then, for every problem $\mathsf{PPEQ}(\mathbf{B}) \in \mathsf{PPEQ}(\mathbb{B})$, there exists a problem $\mathsf{PPEQ}(\mathbf{B}') \in \mathsf{PPEQ}(\mathbb{A})$ such that $\mathsf{PPEQ}(\mathbf{B})$ reduces to $\mathsf{PPEQ}(\mathbf{B}')$, and likewise for $\mathsf{PPCON}(\cdot)$.*

**Proof**. We first treat powers; suppose $\mathbb{B} = \mathbb{A}^k$. Consider a problem $\mathsf{PPEQ}(\mathbf{B}) \in \mathsf{PPEQ}(\mathbb{A}_\mathbf{B})$, and let $\sigma$ denote the signature of $\mathbf{B}$. Let $\sigma'$ be the signature that has the same symbols as $\sigma$, but where the arity of a symbol of $R \in \sigma'$ is $km$, where $m$ is the arity of $R \in \sigma$. Define $\mathbf{B}'$ to be the structure whose relation $R^{\mathbf{B}'}$ contains the tuple $(a_1^1, \ldots, a_1^k, \ldots, a_m^1, \ldots, a_m^k)$ if and only if the tuple $((a_1^1, \ldots, a_1^k), \ldots, (a_m^1, \ldots, a_m^k))$ belongs to the relation $R^\mathbf{B}$. Clearly, we have $\mathsf{PPEQ}(\mathbf{B}') \in \mathsf{PPEQ}(\mathbb{A})$. To reduce an instance $(\phi, \phi')$ of $\mathsf{PPEQ}(\mathbf{B})$ to $\mathsf{PPEQ}(\mathbf{B}')$, we simply replace, in each of $\phi, \phi'$, each variable $v$ with a sequence of $k$ variables $v^1, \ldots, v^k$. It is straightforward to verify that the original instance $(\phi, \phi')$ was a yes instance if and only if the new formulas are. The same reduction applies to $\mathsf{PPCON}(\cdot)$.

In the case that $\mathbb{B}$ is a subalgebra or homomorphic image of $\mathbb{A}$, the result is proved in [6, Proposition 4] for $\mathsf{PPEQ}(\cdot)$, and from the argumentation there it is clear that exactly the same reduction works for $\mathsf{PPCON}(\cdot)$. □

## 3 Unary Type

In this section, we present the first hardness result described in the introduction.

Our proof makes use of the detailed information on tame congruence theory provided in [13] and [17]. This theory associates a *typeset* to a non-trivial finite algebra, which contains one or more of five *types*: (1) the unary type, (2) the affine type, (3) the boolean type, (4) the lattice type, and (5) the semilattice type. By extension, a typeset is associated to each variety, namely, the union of all typesets of finite algebras contained in the variety. A variety is said to *admit* a type if the type is contained in its typeset, and is otherwise said to *omit* the type.

**Theorem 5.** *Let $\mathbf{B}$ be a finite relational structure. If $\mathcal{V}(\mathbb{A}_\mathbf{B})$ admits the unary type, then $\mathsf{PPEQ}(\mathbf{B})$ and $\mathsf{PPCON}(\mathbf{B})$ are $\Pi_2^p$-hard.*

As we now show, modulo the G-Set conjecture, the previous theorem implies a coNP/$\Pi_2^p$-complete dichotomy for the equivalence and containment problems.

**Theorem 6.** *Let* **B** *be a finite relational structure over a finite signature. If the G-Set conjecture holds and* $\mathcal{V}(\mathbb{A}_\mathbf{B})$ *omits the unary type, then both* PPEQ(**B**) *and* PPCON(**B**) *are contained in* coNP.

*Proof.* Let $\mathbf{B}^*$ be obtained from **B** by adding to it all relations of the form $\{b\}$ for $b \in B$. If **B** is a finite relational structure such that the variety generated by $\mathbb{A}_\mathbf{B}$ omits the unary type, then the variety generated by $\mathbb{A}_{\mathbf{B}^*}$ omits the unary type also, and according to the G-Set conjecture, CSP($\mathbf{B}^*$) is in P. From this it follows that PPEQ(**B**) and PPCON(**B**) are in coNP: this is because deciding if $\mathbf{B}, f \models \phi$ for a pp-formula $\phi$ can be viewed as an instance of CSP($\mathbf{B}^*$) when the free variables of $\phi$ are fixed to constants according to $f$. □

In order to prove Theorem 5, we will first embark on a study of varieties omitting the unary type (Section 3.1), and establish the key algebraic lemma (Lemma 1). Then (Section 3.2), we establish the desired hardness result by reducing from the containment problem over a boolean structure having constant polymorphisms, known to be $\Pi_2^p$-complete [6], to the containment problem of interest (Theorem 5).

### 3.1 Algebra

The following algebraic lemma is the key to the hardness result under consideration.

**Lemma 1.** *Let* **B** *be a finite relational structure such that* $\mathcal{V}(\mathbb{A}_\mathbf{B})$ *admits the unary type, and let* $\mathbf{C} = (\{0,1\}, \{C_1, \ldots, C_l\})$ *be a relational structure whose relations contain the constant tuples. Then, there is a finite algebra* $\mathbb{A} \in \mathcal{V}(\mathbb{A}_\mathbf{B})$ *and a finite set*

$$\mathcal{R} = \{D_1, \ldots, D_l\} \cup \{E_1, \ldots, E_k\}$$

*of finitary relations over $A$ (where $k = |A|$), compatible with the operations of* $\mathbb{A}$, *satisfying the following.*

*Let* $\mathbf{A} = (A, \mathcal{R})$. *Let* $\phi(x_1, \ldots, x_m)$ *be a pp-formula on* **C** *with quantified variables* $x_{m+1}, \ldots, x_n$. *Define the pp-formula* $\phi'(x_1, \ldots, x_m)$ *over* **A** *by replacing each atomic formula* $C_i(z_1, \ldots, z_r)$ *in* $\phi$ *by* $D_i(z_1, \ldots, z_r)$, *and conjoining* $E_n(x_1, \ldots, x_n)$ *if* $n \leq k$, *and*

$$\bigwedge_{1 \leq i_1 < \cdots < i_k \leq n} E_k(x_{i_1}, \ldots, x_{i_k}), \tag{1}$$

*otherwise ($n > k$). Then,*

$$(\phi, \psi) \in \mathsf{PPCON}(\mathbf{C}) \text{ if and only if } (\phi', \psi') \in \mathsf{PPCON}(\mathbf{A}). \tag{2}$$

*Proof.* The proof requires a certain amount of the theory of tame congruences and multitraces; for further background, we refer the reader to [13] and [17] respectively.

If $\mathcal{V} = \mathcal{V}(\mathbb{A}_\mathbf{B})$ admits the unary type, then by [13, Theorem 6.17 and Lemma 6.18] there exists a finite algebra $\mathbb{A}$ in $\mathcal{V}$ and a congruence $\alpha$ on $\mathbb{A}$ such that: $\alpha$ covers $0_A$ in Con ($\mathbb{A}$); the type of the congruence pair $(0_A, \alpha)$ is unary; and the $(0_A, \alpha)$-traces are all polynomially equivalent to two-element sets. Fix such an algebra $\mathbb{A}$ and such a congruence $\alpha$ on $\mathbb{A}$, and choose some $(0_A, \alpha)$-minimal set $U$, and some $(0_A, \alpha)$-trace $N = \{0, 1\}$ contained in $U$. In the sequel, $n > 0$.

For an $n$-ary relation $R \subseteq N^n$ that contains the constant tuples, following [13, Definition 6.13], we define the $n$-ary relation $\mathbb{A}(R) \subseteq A^n$ to be the universe of the subalgebra of $\mathbb{A}^n$ generated by $R \cup \{(a, a, \ldots, a) \mid a \in A\}$. We record the following [13, Lemma 6.14(2) and Corollary 5.2(2)].

**Fact 7** *If $M = \{a, b\}$ is a $(0_A, \alpha)$-trace and $p(x)$ is a polynomial of* $\mathbb{A}$ *with $p(0) = a$ and $p(1) = b$, then* $\mathbb{A}(R) \cap M^n = p(R)$. *Moreover, if $(a_1, \ldots, a_n) \in \mathbb{A}(R)$, then $(a_i, a_j) \in \alpha$ for all $1 \leq i \leq j \leq n$ and if $q(x)$ is a unary polynomial of* $\mathbb{A}$, *then* $(q(a_1), \ldots, q(a_n)) \in \mathbb{A}(R)$.

Notice that in particular $\mathbb{A}(R) \cap N^n = R$. We define

$$E_n = \mathbb{A}(N^n),$$

and we observe the following.

**Fact 8** *Let $n > k$. Then $(a_1, \ldots, a_n) \in E_n$ iff $(a_{i_1}, \ldots, a_{i_k}) \in E_k$ for all $1 \leq i_1 < \cdots < i_k \leq n$.*

Thus, for every $n > k$, the formula (1) is a pp-definition over $\mathbf{A}$ of the relation $E_n \subseteq A^n$.

Now let $\mathbf{C} = (\{0, 1\}, \{C_1, \ldots, C_l\})$ be a relational structure whose relations contain the constant tuples. Let $D_i = \mathbb{A}(C_i)$ for $i = 1, \ldots, l$ and let

$$\mathcal{R} = \{D_1, \ldots, D_l\} \cup \{E_1, \ldots, E_k\}.$$

Clearly, $\mathcal{R}$ is a finite set of finitary relations over $A$ compatible with the operations of $\mathbb{A}$. Finally, statement (2) follows immediately from Proposition 10, ending the proof of this Lemma. □

Our proof of the following Proposition appeals to the theory of multitraces [17] and to Facts 7 and 8. A *multitrace* is a subset $T$ of $A$ of the form $f(N, N, \ldots, N)$, where $f$ is some polynomial of $\mathbb{A}$ and $N$ is the $(0_A, \alpha)$-trace from the Lemma. We record some relevant facts about multitraces in the following Theorem.

**Theorem 9.** *(follows from [17, Theorem 3.10]) Let $T$ be a multitrace, say $T = f(N, N, \ldots, N)$ for some $m$-ary polynomial $f$ of $\mathbb{A}$. There is a $p$-ary polynomial $f'(\bar{x})$ of $\mathbb{A}$ for some $p \leq m$ and some unary polynomials (called coordinate maps) $g_i(x)$ of $\mathbb{A}$, for $1 \leq i \leq p$, such that*

- $T = f'(N, N, \ldots, N)$ *and* $g_i(T) \subseteq N$ *for all* $i$,
- *for all* $x_j \in N$, *and all* $i$, $g_i(f'(x_1, \ldots, x_p)) = x_i$,
- *for all* $x \in T$, $x = f'(g_1(x), \ldots, g_p(x))$,
- *the set $N^p$ is in bijective correspondence with $T$ via the map that takes a $p$-tuple $(n_1, \ldots, n_p)$ to $f'(n_1, \ldots, n_p)$.*

**Proposition 10** *If $\phi(x_1, \ldots, x_m)$ is a pp-formula over $\mathbf{C}$, and $\phi'(x_1, \ldots, x_m)$ is the pp-formula over $\mathbf{A}$ as in the statement, then, where $X = \{x_1, \ldots, x_m\}$: For every $g: X \to \{0, 1\}$,*

$$\mathbf{C}, g \models \phi \text{ iff } \mathbf{A}, g \models \phi';$$

*and, for every $g: X \to A$,*

$$\mathbf{A}, g \models \phi' \text{ iff } g \in \mathbb{A}(\{g' \mid \mathbf{C}, g' \models \phi\}).$$

*Proof.* Suppose that $Y = \{x_{m+1}, \ldots, x_n\}$ is the set of quantified variables of $\phi$. For the first equivalence, if $g: X \to \{0, 1\}$ and $\mathbf{C}, g \models \phi$ is witnessed by the elements $b_j \in \{0, 1\}$, $m+1 \leq j \leq n$, then all of these elements are in $N$ and so $(g(x_1), \ldots, g(x_m), b_{m+1}, \ldots, b_n) \in E_n$. If some clause $C_i(z_1, \ldots, z_r)$ of $\phi$ holds for $g$ and the $b_j$'s, then by construction $D_i(z_1, \ldots, z_r)$ also holds for these elements. From this it follows that $\mathbf{A}, g \models \phi'$, using the same witnesses $b_j$, $m + 1 \leq j \leq n$.

Conversely, suppose that $g: X \to \{0, 1\}$ and $\mathbf{A}, g \models \phi'$ is witnessed by the elements $b_j \in A$, $m + 1 \leq j \leq n$. We make use of a unary polynomial $e(x)$ of $\mathbb{A}$ provided by Tame Congruence Theory with the properties that $e(e(x)) = e(x)$ holds on $A$, the range of $e$ contains 0 and 1, and if $a$ is any element of $A$ that is $\alpha$-related to 0 then $e(a) \in \{0, 1\}$. Since each relation $D_i$ that appears as a clause in $\phi'$ is closed under the unary operation $e(x)$, applied coordinatewise (see Fact 7), it follows that the elements $e(b_j)$, $m + 1 \leq j \leq n$ also witness that $\mathbf{A}, g \models \phi'$. Here we use that since the tuple $(g(x_1), \ldots, g(x_m), b_{m+1}, \ldots, b_n)$ belongs to $E_n$, then so does the tuple $(g(x_1), \ldots, g(x_m), e(b_{m+1}), \ldots, e(b_n))$. We also note that this implies that the $e(b_j)$'s are all $\alpha$-related to 0 and so belong to $\{0, 1\}$.

Finally, Fact 7 provides that $D_i \cap \{0, 1\}^p = C_i$, where $p$ is the arity of $C_i$, and from this it follows that the elements $e(b_j)$, $m + 1 \leq j \leq n$, also witness that $\mathbf{C}, g \models \phi$. Thus the first equivalence has been established.

For the second equivalence, we can apply the first equivalence to conclude that for any $g\colon X \to A$, $g$ belongs to $\mathbb{A}(\{g'\colon X \to \{0,1\} \mid \mathbf{C}, g' \models \phi\})$ if and only if it belongs to $\mathbb{A}(\{g'\colon X \to \{0,1\} \mid \mathbf{A}, g' \models \phi'\})$. So it will suffice to prove that

$$\mathbb{A}(\{g'\colon X \to \{0,1\} \mid \mathbf{A}, g' \models \phi'\}) = \{g\colon X \to A \mid \mathbf{A}, g \models \phi'\}.$$

to complete the proof.

The containment of the left hand side of this derived equality in the set $\{g\colon X \to A \mid \mathbf{A}, g \models \phi'\}$ follows after noting that both sets in the equality are subuniverses of $\mathbb{A}^m$ and that all of the generators of the subuniverse on the left hand side are contained in the subuniverse on the right hand side. Here we use the fact that each relation $D_i$ of $\mathbf{A}$ contains all constant tuples and so every constant $m$-tuple over $A$ is a member of the subuniverse on the right hand side.

For the remaining containment, assume that $g\colon X \to A$ is such that $\mathbf{A}, g \models \phi'$ and suppose that the elements $b_i \in A$, $m+1 \leq i \leq n$ witness this. Since $(g(x_1), \ldots, g(x_m), b_{m+1}, \ldots, b_n) \in E_n$ then there is some multitrace $T$ of $\mathbb{A}$ that contains all of these elements. By Theorem 9, it follows that for some $p > 0$, there is some $p$-ary polynomial $f'$ and coordinate maps $g_i(x)$, $1 \leq i \leq p$, that satisfy the properties stated in that theorem. In particular, $T = f'(N, N, \ldots, N)$ and for all $c \in T$, $c = f'(g_1(c), \ldots, g_p(c))$.

From Fact 7 it follows that for any relation $D_i$ of $\mathbf{A}$, with $D_i$ of arity $q$, if $c_j \in T$, for $1 \leq j \leq q$, and $(c_1, \ldots, c_q) \in D_i$, then

$$(g_j(c_1), \ldots, g_j(c_q)) \in (D_i \cap \{0,1\}^q) = C_i,$$

for any $j$. This is because for each coordinate map $g_j$, we have that $g_j(T) \subseteq \{0,1\}$. Extending this to our pp-formula $\phi'$, it follows that not only is $(g_j(g(x_1)), \ldots, g_j(g(x_m)))$ a solution, witnessed by $g_j(b_l)$, $m+1 \leq l \leq n$, but it is also a member of $\{0,1\}^m$. So, each of $g_j \circ g$ belongs to the generating set of the relation $\mathbb{A}(\{g'\colon X \to \{0,1\} \mid \mathbf{A}, g' \models \phi'\})$.

Since for each $j$, $g(x_j) = f'(g_1(g(x_j)), \ldots, g_p(g(x_j)))$ and $\mathbb{A}(\{g'\colon X \to \{0,1\} \mid \mathbf{A}, g' \models \phi'\})$ is closed under $f'$, applied coordinatewise, we conclude that $g$ is also a member of this relation, as required. □

### 3.2 Reduction

We now prove Theorem 5.

*Proof (Theorem 5).* By Theorem 4 and Lemma 4 from [6] it follows that there exists a Boolean relational structure $\mathbf{C} = (\{0,1\}, \{C_1, \ldots, C_l\})$, whose relations contain the constant tuples, such that $\mathsf{PPCON}(\mathbf{C})$ is $\Pi_2^p$-complete. Let $\mathbf{A} = (A, \{D_1, \ldots, D_l, E_1, \ldots, E_k\})$ be the finite relational structure defined in terms of Lemma 1 over the universe $A$ of the finite algebra $\mathbb{A} \in \mathcal{V}(\mathbb{A}_\mathbf{B})$. We describe a reduction from $\mathsf{PPCON}(\mathbf{C})$ to $\mathsf{PPCON}(\mathbf{A})$. Notice that $\mathbf{A}$ is such that $\mathsf{PPCON}(\mathbf{A}) \in \mathsf{PPCON}(\mathbb{A})$, thus $\mathsf{PPCON}(\mathbf{B})$ is $\Pi_2^p$-hard by Theorem 4; $\Pi_2^p$-hardness of $\mathsf{PPEQ}(\mathbf{B})$ follows from Proposition 1.

The reduction, given an instance $(\phi, \psi)$ of $\mathsf{PPCON}(\mathbf{C})$, returns an instance $(\phi', \psi')$ of $\mathsf{PPCON}(\mathbf{A})$, where $\phi'$ and $\psi'$ are defined from $\phi$ and $\psi$ as in Lemma 1. The reduction is therefore correct, and polynomial-time since the pp-definition of $E_n$ over the relations in $\{E_1, \ldots, E_k\}$ has size polynomial in $n$ for every $n \geq 1$. □

## 4 Non-Congruence Modularity

In this section, we present the second hardness result described in the introduction.

An algebra $\mathbb{A}$ is said to be congruence modular, if its lattice of congruences satisfies the modular law:

$$\forall x \forall y \forall z \left[ x \leq y \to x \vee (y \wedge z) = y \wedge (x \vee z) \right].$$

A variety is said to be congruence modular if all of its members are congruence modular.

**Theorem 11.** *Let* **B** *be a finite relational structure. If* $\mathcal{V}(\mathbb{A}_\mathbf{B})$ *is not congruence modular, then* PPEQ(**B**) *and* PPCON(**B**) *are coNP-hard.*

The previous theorem implies, modulo the Edinburgh conjecture, a P/coNP-hard dichotomy for the equivalence and containment problems.

**Theorem 12.** *Let* **B** *be a finite relational structure over a finite signature. If the Edinburgh conjecture holds and* $\mathcal{V}(\mathbb{A}_\mathbf{B})$ *is congruence modular, then* PPEQ(**B**) *and* PPCON(**B**) *are in P.*

*Proof.* By the Edinburgh conjecture, the congruence modularity of $\mathcal{V}(\mathbb{A}_\mathbf{B})$ implies that $\mathbb{A}_\mathbf{B}$ has few subpowers. Then, we have from [6, Theorem 7] that PPEQ(**B**) and PPCON(**B**) are in P. □

In order to prove Theorem 11, we will first embark on a study of varieties that fail to be congruence modular (Section 4.1), and establish the key algebraic lemma (Lemma 2). We will then give a sequence of reductions to establish the desired hardness result (Section 4.2): we first reduce from the problem of deciding whether a DNF is a tautology, a known coNP-complete problem, to a certain comparison problem over lattices; we then reduce to a certain entailment problem on 2-sorted relational structures derived from structures which we call "pentagons"; finally, we reduce from this entailment problem to the containment problem (Theorem 11).

### 4.1 Algebra

Let $A$ be a set. For binary relations $\theta, \theta'$ on $A$, the relational product $\theta \circ \theta'$ is the binary relation defined by $\{(a,b) \mid \exists c.(a,c) \in \theta, (c,b) \in \theta'\}$. We use $\theta^k$ to denote the $k$-fold relational product of $\theta$ with itself. We let $\mathrm{Eq}(A)$ denote the complete lattice of equivalence relations on $A$, and we let $0_A = \{(a,a) \mid a \in A\}$ and $1_A = A^2$ denote the bottom and top elements of $\mathrm{Eq}(A)$, respectively.

**Proposition 13** *Let $A$ be a set such that $|A| = m$. Let $\theta_1, \ldots, \theta_k \in \mathrm{Eq}(A)$. It holds that $\theta_1 \vee \cdots \vee \theta_k = (\theta_1 \circ \cdots \circ \theta_k)^m$.*

*Proof.* It is immediate that $(\theta_1 \circ \cdots \circ \theta_k)^m \subseteq \theta_1 \vee \cdots \vee \theta_k$.

If $(a,b) \in \theta_1 \vee \cdots \vee \theta_k$, then there exist $e_0, \ldots, e_l \in A$ with $l \leq m$ such that $e_0 = a$, $(e_{i-1}, e_i) \in \theta_j$ for some $j \in \{1, \ldots, k\}$, and $e_l = b$. By the reflexivity of the $\theta_i$, for all $j \in \{1, \ldots, k\}$ and $c, d \in A$, if $(c,d) \in \theta_j$, then $(c,d) \in \theta_1 \circ \cdots \circ \theta_k$. Thus, $(a,b) \in (\theta_1 \circ \cdots \circ \theta_k)^l \subseteq (\theta_1 \circ \cdots \circ \theta_k)^m$. □

A *pentagon* is a structure **P** over the signature $\{\alpha, \beta, \gamma\}$ containing three binary relation symbols such that $\alpha^\mathbf{P}$, $\beta^\mathbf{P}$, and $\gamma^\mathbf{P}$ are equivalence relations on $P$, and the following conditions hold in $\mathrm{Eq}(P)$:

- $\alpha^\mathbf{P} \leq \beta^\mathbf{P}$,
- $\beta^\mathbf{P} \wedge \gamma^\mathbf{P} = 0_P$,
- $\beta^\mathbf{P} \circ \gamma^\mathbf{P} = 1_P$, and
- $\alpha^\mathbf{P} \vee \gamma^\mathbf{P} = 1_P$.

We remark that in the sequence of reductions that we give, we do not make explicit use of the last item in the definition of a pentagon.

A pentagon $\mathbf{P} = (P, \alpha^\mathbf{P}, \beta^\mathbf{P}, \gamma^\mathbf{P})$ can be naturally decomposed as a direct product $P = B \times C$ in such a way that $\beta^\mathbf{P}$ and $\gamma^\mathbf{P}$ are the kernels of the projections of $P$ onto $B$ and $C$, respectively. Each element $b \in B$ induces, via $\alpha^\mathbf{P}$, an equivalence relation $\alpha_b$ on $C$, namely

$$\alpha_b = \{(c,c') \mid ((b,c),(b,c')) \in \alpha^\mathbf{P}\} \in \mathrm{Eq}(C). \tag{3}$$

In associating together two elements $b, b' \in B$ when $\alpha_b = \alpha_{b'}$, one naturally obtains a partition of $B$ into $l \geq 1$ non-empty blocks $B_1, \ldots, B_l$ and equivalence relations $\alpha_1, \ldots, \alpha_l$ on $C$ such that for all $b \in B$, it holds that $\alpha_i = \alpha_b$ if and only if $b \in B_i$. We say that a pentagon is *interesting* if the sequence $\alpha_1, \ldots, \alpha_l$ contains equivalence relations $\alpha_j$ and $\alpha_k$ such that $\alpha_j < \alpha_k$ holds in $\mathrm{Eq}(C)$; we say that a set of pentagons is *interesting* if it contains an interesting pentagon.

Let **B** be a finite relational structure. If $\mathcal{V} = \mathcal{V}(\mathbb{A}_\mathbf{B})$ is not congruence modular, then this can be witnessed in the congruence lattice of the 4-generated free algebra in $\mathcal{V}$. More precisely, let $\mathbb{F}_4$ be the $\mathcal{V}$-free algebra freely generated by **a**, **b**, **c**, and **d** and let $\alpha^*$, $\beta$, and $\gamma$ be the congruences of $\mathbb{F}_4$ generated by $\{(\mathbf{a},\mathbf{b})\}$, $\{(\mathbf{a},\mathbf{b}),(\mathbf{c},\mathbf{d})\}$, and $\{(\mathbf{a},\mathbf{c}),(\mathbf{b},\mathbf{d})\}$ respectively. If we set $\alpha = \alpha^* \vee (\beta \wedge \gamma)$ then it follows ([11]) that $\mathcal{V}$ will fail to be congruence modular if and only if $\alpha < \beta$ and that in this case, the three congruences $\alpha$, $\beta$, and $\gamma$ provide a witness to the failure of the modular law in the congruence lattice of $\mathbb{F}_4$.

By working over a suitable quotient of the algebra $\mathbb{F}_4$, we establish the following Lemma.

**Lemma 2.** *Let **B** be a finite relational structure such that $\mathcal{V}(\mathbb{A}_\mathbf{B})$ is not congruence modular. There is a finite algebra $\mathbb{A} \in \mathcal{V}(\mathbb{A}_\mathbf{B})$ having congruences $\alpha$, $\beta$, and $\gamma$ such that $\alpha < \beta$, $\gamma \wedge \beta = 0_A$, and $\alpha \vee \gamma = \beta \vee \gamma$. Furthermore, there exists a finite interesting set of pentagons $\mathcal{P}$, and a finite set $\mathcal{D}$ of finitary relations over $A$ compatible with the operations of $\mathbb{A}$ such that:*

(i) *If $\mathbf{P} = (P, \alpha^\mathbf{P}, \beta^\mathbf{P}, \gamma^\mathbf{P}) \in \mathcal{P}$, then $P \subseteq A$ and $\alpha^\mathbf{P} = \alpha \cap P^2$, $\beta^\mathbf{P} = \beta \cap P^2$, and $\gamma^\mathbf{P} = \gamma \cap P^2$.*
(ii) *For every $k \geq 1$, there exists a $k$-ary relation $D_k$ on $A$, such that $D_k$ has a pp-definition over the relations in $\mathcal{D}$ with size polynomial in $k$, and such that for all $a_1, \ldots, a_k \in A$, $(a_1, \ldots, a_k) \in D_k$, if and only if $a_1, \ldots, a_k \in P$ for some $\mathbf{P} = (P, \alpha^\mathbf{P}, \beta^\mathbf{P}, \gamma^\mathbf{P}) \in \mathcal{P}$.*

*Proof.* As noted earlier in this subsection, since $\mathcal{V} = \mathcal{V}(\mathbb{A}_\mathbf{B})$ is not congruence modular, then the congruences $\alpha$, $\beta$, and $\gamma$ of $\mathbb{F}_4$ provide a witness to this. In general, $\beta \wedge \gamma$ will be some non-trivial congruence of $\mathbb{F}_4$ and so we will eventually take the quotient of $\mathbb{F}_4$ by $\beta \wedge \gamma$ to obtain the desired algebra $\mathbb{A}$ and we will replace $\alpha$, $\beta$, and $\gamma$ by the congruences $\alpha/(\beta \wedge \gamma)$, $\beta/(\beta \wedge \gamma)$, and $\gamma/(\beta \wedge \gamma)$ in the congruence lattice of $\mathbb{A}$.

Let $\delta = \beta \vee \gamma$, a congruence of $\mathbb{F}_4$, and for each $u \in F_4$, let $[u]$ denote the $\delta$-class that contains $u$.

*Claim.*
1. For each term $t(x_1, \ldots, x_m)$ of $\mathbb{F}_4$ the set $t([\mathbf{a}]^m)$ is contained in a single $\delta$-class, namely the class that contains the element $t(\mathbf{a}, \ldots, \mathbf{a})$ and each $\delta$-class $[u]$ is a union of sets of this form.
2. For each $u \in F_4$, there is a unary term $t_u(x)$ such that $[u] = [t_u(\mathbf{a})]$. The term $t_u$ is unique up to equality in $\mathcal{V}$. Consequently $(u,v) \in \delta$ if and only if the equation $t_u(x) = t_v(x)$ holds in $\mathcal{V}$.
3. The class $[\mathbf{a}]$ is equal to the set of all elements of the form $t(\mathbf{a}, \mathbf{b}, \mathbf{c}, \mathbf{d})$ where $t$ is an idempotent term of $\mathcal{V}$. If $u = t(\mathbf{a}, \mathbf{b}, \mathbf{c}, \mathbf{d})$ is in $[\mathbf{a}]$ then $t$ is an idempotent term.

*Proof (of Claim).* Since all elements of the set $[\mathbf{a}]$ are $\delta$-related and $t$ is compatible with $\delta$, then any two elements from $t([\mathbf{a}]^m)$ are $\delta$-related, and in fact $\delta$-related to the element $t(\mathbf{a}, \ldots, \mathbf{a})$. On the other hand, if $u \in F_4$, then for some term $t$, $u = t(\mathbf{a}, \mathbf{b}, \mathbf{c}, \mathbf{d})$, since $\{\mathbf{a}, \mathbf{b}, \mathbf{c}, \mathbf{d}\}$ generates $\mathbb{F}_4$, and so $u$ is contained in the set $t([\mathbf{a}]^4)$. If $u \in F_4$ is equal to $t(\mathbf{a}, \mathbf{b}, \mathbf{c}, \mathbf{d})$ for some term $t(x, y, z, w)$, then let $t_u(x) = t(x, x, x, x)$. Since the free generators of $\mathbb{F}_4$ are all $\delta$-related, it follows that the element $t_u(\mathbf{a})$ is $\delta$-related to $u$ and so $[u] = [t_u(\mathbf{a})]$. If $(u, v) \in \delta$ then $t_u(\mathbf{a})$ is $\delta$-related to $t_v(\mathbf{a})$ and it follows from Lemma 3.6 of [16] that in fact $t_u(\mathbf{a}) = t_v(\mathbf{a})$ and so the equation $t_u(x) = t_v(x)$ holds in $\mathcal{V}$. The third part of this claim also follows directly from the same Lemma. $\square$

For a term $t(x_1, \ldots, x_m)$ of $\mathbb{F}_4$, let $\hat{S}[t] = t([\mathbf{a}]^m)$ and for any congruence $\theta$ of $\mathbb{F}_4$, let $\theta_t = \theta \cap (\hat{S}[t])^2$, the restriction of $\theta$ to the set $\hat{S}[t]$.

*Claim.* Let $t(x_1, \ldots, x_m)$ be a term of $\mathbb{F}_4$ and let $S = \hat{S}[t]$.

1. If $t$ is idempotent, then $S = [\mathbf{a}]$ and $\alpha_t < \beta_t$. In general, $S$ will be a proper subset of the $\delta$-class that contains it.
2. $\alpha_t \vee \gamma_t = 1_S$,
3. $\beta_t$ and $\gamma_t$ permute,
4. at least one $\alpha_t$ class is equal to a $\beta_t$ class,

*Proof (of Claim).* If $t$ is idempotent and $u \in [\mathbf{a}]$ then $u = t(u, u, \ldots, u) \in S$. On the other hand, any element of the form $t(u_1, u_2, \ldots, u_m)$ with $u_i \in [\mathbf{a}]$ is $\delta$-related to $t(u_1, u_1, \ldots, u_1) = u_1 \in [\mathbf{a}]$ and so belongs to $[\mathbf{a}]$. Since the pair $(\mathbf{c}, \mathbf{d}) \in \beta \setminus \alpha$ (otherwise $\alpha = \beta$) it follows that $\alpha_t < \beta_t$ when $t$ is idempotent.

If $u, v \in S$ then for some $u_i$ and $v_i$ from $[\mathbf{a}]$ we have $u = t(u_1, u_2, \ldots, u_m)$ and $v = t(v_1, v_2, \ldots, v_m)$. Since $(u_i, v_i) \in \delta = \alpha \vee \gamma$ for $1 \le i \le m$ it follows that $(u, v) \in \alpha_t \vee \gamma_t$, i.e., $\alpha_t \vee \gamma_t = 1_S$.

We first show that $\beta_t$ and $\gamma_t$ permute when $t$ is $x$ (or equivalently, when $S = [\mathbf{a}]$). It is immediate, that if $\mathbf{x}, \mathbf{y} \in \{\mathbf{a}, \mathbf{b}, \mathbf{c}, \mathbf{d}\}$ then $(\mathbf{x}, \mathbf{y}) \in \beta_t \circ \gamma_t$. From this we can conclude that if $u = s(\mathbf{a}, \mathbf{b}, \mathbf{c}, \mathbf{d})$ is in $[\mathbf{a}]$ for some (idempotent) term $s$, then $(u, \mathbf{y}) \in \beta_t \circ \gamma_t$ for any $\mathbf{y} \in \{\mathbf{a}, \mathbf{b}, \mathbf{c}, \mathbf{d}\}$. If $v = s'(\mathbf{a}, \mathbf{b}, \mathbf{c}, \mathbf{d})$, for some idempotent term $s'$, is any other member of $S$ then

$$u = s'(u, u, u, u)(\beta_t \circ \gamma_t)s'(\mathbf{a}, \mathbf{b}, \mathbf{c}, \mathbf{d}) = v,$$

thereby showing that $\beta_t \circ \gamma_t = 1_S$, or that $\beta_t$ and $\gamma_t$ permute, in this case. For an arbitrary $t$ the same conclusion can be reached by noting that if $u, v \in S$, say $u = t(u_1, u_2, \ldots, u_m)$ and $v = t(v_1, v_2, \ldots, v_m)$ for some $u_i, v_i \in [\mathbf{a}]$, then since $(u_i, v_i) \in \beta_x \circ \gamma_x$, it follows that $(u, v) \in \beta_t \circ \gamma_t$.

We claim that the $\alpha$ class that contains $\mathbf{a}$ is equal to the $\beta$ class that contains $\mathbf{a}$. It will suffice to show that if $(\mathbf{a}, u) \in \beta$, then $(\mathbf{a}, u) \in \alpha^* \vee (\beta \wedge \gamma)$. Since $u \in [\mathbf{a}]$ then there is some idempotent term $s$ such that $u = s(\mathbf{a}, \mathbf{b}, \mathbf{c}, \mathbf{d})$. The element $s(\mathbf{a}, \mathbf{b}, \mathbf{a}, \mathbf{b})$ is $\gamma$-related to $u$ and $\alpha^*$-related to $\mathbf{a}$ and hence is also $\beta$-related to $u$ (since $\alpha^* < \beta$ and $(\mathbf{a}, u) \in \beta$). But then we have that $\mathbf{a}\alpha^* s(\mathbf{a}, \mathbf{b}, \mathbf{a}, \mathbf{b})(\beta \wedge \gamma)u$, as required.

For arbitrary $t$, let $u = t(\mathbf{a}, \mathbf{a}, \ldots, \mathbf{a}) \in S$. We claim that every element $v$ of $S$ that is $\beta_t$-related to $u$ is actually $\alpha_t$-related to $u$. If $v = t(v_1, v_2, \ldots, v_m)$ for some $v_i \in [\mathbf{a}]$ then from the previous argument, we can find, for each $i$, elements $w_i \in [\mathbf{a}]$ such that $(\mathbf{a}, w_i) \in \alpha$ and $(w_i, v_i) \in \gamma$ (since $\beta_x \circ \gamma_x = 1_{[\mathbf{a}]}$). The element $w = t(w_1, w_2, \ldots, w_m)$ is in $S$ and is $\alpha_t$-related to $u$. We also have that $(v, w) \in \gamma_t$ and so in fact $(v, w) \in \beta_t \wedge \gamma_t \le \alpha_t$. Thus $(u, v) \in \alpha_t$, as claimed. □

For each $k \ge 1$, let $\hat{D}_k$ be the subuniverse of $\mathbb{F}_4^k$ generated by $[\mathbf{a}]^k$. We establish the following for this set of relations over $F_4$. Let $N = |F_4|$.

*Claim.* 1. For $k > N$, the relation $\hat{D}_k(x_1, \ldots, x_k)$ is equal to

$$\bigcap_{\{1 \le i_1 < i_2 < \cdots < i_N \le k\}} \hat{D}_N(x_{i_1}, \ldots, x_{i_N}).$$

Thus, $\hat{D}_k$ has a pp-definition over $\hat{D}_N$ of size polynomial in $k$.

2. For all $k$ and all $a_1, \ldots, a_k \in F_4$, $(a_1, \ldots, a_k) \in \hat{D}_k$ if and only if $\{a_1, \ldots, a_k\} \subseteq \hat{S}[t]$ for some term $t$ of $\mathbb{F}_4$.

*Proof (of Claim).* Since the projection of $\hat{D}_k$ onto any subset of $n$ variables, with $n \le k$ is equal to the relation $\hat{D}_n$, it follows that any $k$-tuple in $\hat{D}_k$ is contained in the displayed intersection. Conversely, if $(a_1, \ldots, a_k)$ is some $k$-tuple that is in the displayed intersection, then, since $|F_4| = N$, there is some sequence $1 \le i_1 < i_2 < \cdots < i_N \le k$ such that for all $j \le k$, $a_j = a_{i_l}$ for some $l \le N$. Since the tuple $(a_{i_1}, \ldots, a_{i_N}) \in \hat{D}_N$, then there is some term $t(x_1, \ldots, x_m)$ and $N$-tuples $\mathbf{u}_i = (u_i^1, \ldots, u_i^N) \in [\mathbf{a}]^N$, for $1 \le i \le m$ such that $(a_{i_1}, \ldots, a_{i_N}) = t(\mathbf{u}_1, \ldots, \mathbf{u}_m)$ (with the operation $t$ applied coordinate-wise to the $\mathbf{u}_i$). By the choice of the coordinates $i_j$, the $N$-tuples $\mathbf{u}_i$ can be extended to $k$-tuples $\mathbf{v}_i$ from $[\mathbf{a}]^k$ so that $(a_1, \ldots, a_k) = t(\mathbf{v}_1, \ldots, \mathbf{v}_m)$, thereby showing that $(a_1, \ldots, a_k)$ is in $\hat{D}_k$.

A $k$-tuple $(a_1, \ldots, a_k)$ will be in $\hat{D}_k$ if and only if it can be written as $t(\mathbf{v}_1, \ldots, \mathbf{v}_m)$ for some term $t$ of $\mathbb{F}_4$ and some $k$-tuples $\mathbf{v}_i$ from $[\mathbf{a}]^k$ (since $\hat{D}_k$ is the subuniverse of $\mathbb{F}_4^k$ generated by $[\mathbf{a}]^k$) if and only if $\{a_1, \ldots, a_k\}$ is a subset of $t([\mathbf{a}]^m)$ for some term $t$ if and only if $\{a_1, \ldots, a_k\} \subseteq \hat{S}[t]$ for some term $t$. □

Using the three previous claims we are in a position to conclude the proof of this Lemma. Since the set $F_4$ is finite, then we can select a finite number of terms $t_i$ of $\mathbb{F}_4$, $1 \le i \le p$, with $t_1 = x$, so that for any term $t$ of $\mathbb{F}_4$, we have $\hat{S}[t] = \hat{S}[t_i]$ for some $i \leq p$. For each $i \leq p$, let $\hat{\mathbf{P}}_i$ be the structure with universe $\hat{S}[t_i]$ and relations $\alpha_{t_i}$, $\beta_{t_i}$, and $\gamma_{t_i}$.

If we consider the natural map from $\mathbb{F}_4$ to $\mathbb{A} = \mathbb{F}_4/(\beta \wedge \gamma)$, then each of the relations $\hat{D}_k$ maps to a relation $D_k$ and each of the structures $\hat{\mathbf{P}}_i$ maps to a structure $\mathbf{P}_i$ over $\mathbb{A}$. If we use $\alpha$, $\beta$, and $\gamma$ to denote the congruences $\alpha/(\beta \wedge \gamma)$, $\beta/(\beta \wedge \gamma)$, and $\gamma/(\beta \wedge \gamma)$, respectively, over $\mathbb{A}$, then the Lemma follows, where $\mathcal{P}$ is the set of $\mathbf{P}_i$, $1 \leq i \leq p$, and $\mathcal{D}$ is the set of relations $D_k$ for $1 \leq k \leq N$. We note that the pentagon $\mathbf{P}_1$ is interesting, since, by the second claim, the $\alpha$-class that contains $\mathbf{a}$ is equal to the $\beta$-class that contains $\mathbf{a}$, and the $\alpha$-class that contains $\mathbf{c}$ is a proper subset of the $\beta$-class that contains this element. □

### 4.2 Reductions

**From DNF-TAUTOLOGY to Lattice Inequality.** A propositional formula is in disjunctive normal form (DNF) if it is a finitary disjunction ($\vee$) of finitary conjunctions ($\wedge$) of literals; a literal is a variable, $x$, or the negation of a variable, $\bar{x}$. The following problem is well-known to be coNP-complete.

**Problem:** DNF-TAUTOLOGY
**Instance:** A propositional formula $\phi$ in DNF.
**Question:** Is $\phi$ a tautology?

A *lattice term* $t$ is an algebraic term over finitary joins and meets. The *depth* of $t$ is the height of its syntactic tree. Let $L = (L, \wedge, \vee)$ be a lattice. We say that $L$ is *nontrivial* if $|L| > 1$. Let $S \subseteq L$. Relative to a set of variables $X$, an $S$-*assignment* is a map $f\colon X \to S$.

For a set of lattices $\mathcal{L}$, we define the following computational problem.

**Problem:** DEPTH-4-TERM-INEQ($\mathcal{L}$)
**Instance:** A pair $(t, t')$ of lattice terms of depth $\leq 4$.
**Question:** Does $t \leq t'$ hold in all lattices $L \in \mathcal{L}$?

Hunt, Rosenkrantz, and Bloniarz [15] established the following coNP-hardness result for this problem.

**Theorem 14.** *[15] Let $\mathcal{L}$ be a finite set of finite lattices containing a nontrivial lattice. The problem* DEPTH-4-TERM-INEQ($\mathcal{L}$) *is coNP-hard via a polynomial-time many-one reduction $f$ from* DNF-TAUTOLOGY *satisfying the condition: if $(t, t')$ is a no instance in the image of $f$, then for any lattice $K \in \mathcal{L}$ having elements $a, a' \in K$ with $a < a'$, there is an $\{a, a'\}$-assignment witnessing $t \not\leq t'$ in $K$.*

**From Lattice Inequality to Pentagon Entailment.** To each pentagon $\mathbf{P}$, we associate a 2-sorted structure $\mathbf{P}_2$ having $B$ and $C$ as first and second universe, respectively, where $P = B \times C$ according to the decomposition of $P$ provided by the equivalence relations $\beta^{\mathbf{P}}$ and $\gamma^{\mathbf{P}}$. The structure $\mathbf{P}_2$ is over signature $\{R\}$ and has

$$R^{\mathbf{P}_2} = \{(b, c, c') \in B \times C \times C \mid b \in B_i \Rightarrow (c, c') \in \alpha_i\}. \tag{4}$$

We will be interested in sorted pp-formulas over the signature $\{R\}$. Such formulas are required to have a sort (1 or 2) associated with each variable; the permitted atomic formulas are equality between variables of the same sort and predicate applications having the form $R(x, y, y')$ where $x$ has sort 1, and $y$ and $y'$ have sort 2. We define the following computational problem for each set $\mathcal{P}$ of pentagons.

**Problem:** 2-PENTAGON-ENTAILMENT($\mathcal{P}$)
**Instance:** A pair $(\phi, \psi)$ of sorted pp-formulas over the signature $\{R\}$ having the same free variables for each sort.
**Question:** Does $\phi \models \psi$ over all structures $\mathbf{P}_2$ with $\mathbf{P} \in \mathcal{P}$?

**Theorem 15.** *Let $\mathcal{P}$ be a finite set of finite pentagons containing an interesting pentagon. The problem 2-PENTAGON-ENTAILMENT($\mathcal{P}$) is coNP-hard via a polynomial-time reduction $f$ from DNF-TAUTOLOGY satisfying the condition: if $(\phi, \psi)$ is a no instance in the image of $f$, then $\phi \not\models \psi$ is witnessed over $\mathbf{P}_2$ for any interesting pentagon $\mathbf{P} \in \mathcal{P}$.*

*Proof.* For a pentagon $\mathbf{P} \in \mathcal{P}$, let $P = B \times C$ be its decomposition, and let $\alpha_1, \ldots, \alpha_l$ be the equivalence relations on $C$ associated to $\mathbf{P}$. Let $K_\mathbf{P}$ denote the sublattice of Eq($C$) generated by $\alpha_1, \ldots, \alpha_l$. We define $\mathcal{L} = \{K_\mathbf{P} \mid \mathbf{P} \in \mathcal{P}\}$. Notice that $\mathcal{L}$ contains a nontrivial lattice, since there exists an interesting pentagon in $\mathcal{P}$. Hence, the problem DEPTH-4-TERM-INEQ($\mathcal{L}$) is coNP-hard by Theorem 14; let $r$ denote the reduction given by this theorem.

Let $t(\mathbf{x})$ be a lattice term of depth less than or equal to 4 with $\mathbf{x} = (x_1, \ldots, x_n)$. By induction on the structure of $t$, we show how to construct a pp-formula $\phi_t(\mathbf{x}, y, y')$, where the variables of $\mathbf{x}$ are of sort 1 and the variables $y, y'$ are of sort 2. This translation has the property (*): for all $b_1, \ldots, b_n \in B$ and for all $c, c' \in C$, $\phi_t(b_1, \ldots, b_n, c, c')$ holds in $\mathbf{P}_2$ if and only if $(c, c')$ is in the equivalence relation given by $t^{K_\mathbf{P}}(\alpha_{b_1}, \ldots, \alpha_{b_n})$.

- If $t = x_i$, then $\phi_t(\mathbf{x}, y_1, y_2) = R(x_i, y_1, y_2)$. In this case, property (*) is straightforwardly verified from the definition of $R^{\mathbf{P}_2}$.
- If $t = t_1 \wedge \cdots \wedge t_k$, then $\phi_t(\mathbf{x}, y_1, y_2) = \phi_{t_1}(\mathbf{x}, y_1, y_2) \wedge \cdots \wedge \phi_{t_k}(\mathbf{x}, y_1, y_2)$. In this case, property (*) is straightforward to verify.
- If $t = t_1 \vee \cdots \vee t_k$, we reason as follows. Let

$$m = \max\{|C| \mid \mathbf{P} \in \mathcal{P}, P = B \times C\}. \quad (5)$$

Let $z_{0,k}$ and $z_{i,1}, \ldots, z_{i,k}$ for $i = 1, \ldots, m$ be variables such that $z_{0,k} = y_1$, $z_{m,k} = y_2$, and $z_{i,j}$ is a fresh variable of sort 2 otherwise. Then, $\phi_t(\mathbf{x}, y_1, y_2)$ is the pp-formula obtained by existentially quantifying the fresh variables $z_{i,j}$ before the conjunction

$$\bigwedge_{i=1}^{m} \left( \phi_{t_1}(\mathbf{x}, z_{i-1,k}, z_{i,1}) \wedge \bigwedge_{j=2}^{k} \left( \phi_{t_j}(\mathbf{x}, z_{i,j-1}, z_{i,j}) \right) \right).$$

In this case, property (*) follows from Proposition 13.

The desired reduction is the composition of the reduction $r$ given by Theorem 14 with the mapping $(t, t') \to (\phi_t, \phi_{t'})$. We verify that the reduction is correct. Suppose that $(t, t')$ is a yes instance in the image of $r$. Then, $t \leq t'$ holds in all lattices $K_\mathbf{P}$ with $\mathbf{P} \in \mathcal{P}$. It follows immediately from property (*) that $\phi_t \models \phi_{t'}$ over all pentagons $\mathbf{P}_2$ with $\mathbf{P} \in \mathcal{P}$. Suppose now that $(t(x_1, \ldots, x_n), t'(x_1, \ldots, x_n))$ is a no instance in the image of $r$. Let $\mathbf{P} \in \mathcal{P}$ be an interesting pentagon. There exist $b_1, b_2 \in B$ with $\alpha_{b_1} < \alpha_{b_2}$ in $K_\mathbf{P}$. By Theorem 14, there exists an $\{\alpha_{b_1}, \alpha_{b_2}\}$-assignment $g$ defined on $\{x_1, \ldots, x_n\}$ such that $t(g) \not\leq t'(g)$ in $K_\mathbf{P}$. Let $h$ be the $\{b_1, b_2\}$-assignment on $\{x_1, \ldots, x_n\}$ naturally induced by $g$, and let $(c, c') \in C \times C$ be such that $(c, c') \in t(g) \setminus t'(g)$. From property (*), we have that $\phi_t \not\models \phi_{t'}$ is witnessed over $\mathbf{P}_2$ by the assignment $h, (c, c')$.

It remains to show that the translation $t \to \phi_t$ can be computed in polynomial time. Let $s(t)$ denote the size $|\phi_t|$ of a term $t$. We prove that for a bounded-depth term $t$, the size $s(t)$ is polynomial in $|t|$, the size of $t$, which suffices. By inspection of the translation $t \to \phi_t$, there exist natural numbers $L, B, E$ that are polynomial in $|t|$ such that

- for a term $t = x_i$, it holds that $s(t) \leq L$.
- for a term $t = t_1 \wedge \cdots \wedge t_k$ or a term $t = t_1 \vee \cdots \vee t_k$, it holds that $s(t) \leq B(s(t_1) + \cdots + s(t_k)) + E$.

Now, we define the function $u$ recursively as follows:

- $u(0, n) = Ln$.
- $u(d+1, n) = Bnu(d, n) + E$.

We prove the following claim: for all terms $t$, it holds that $s(t) \leq u(d, n)$, where $d$ is the depth of $t$ and $n$ is the number of leaves (that is, the number of variable occurrences) of $t$. This suffices, as for each fixed $d$, the function $u$ can be viewed as a polynomial in $L$, $B$, $E$, and $n$.

For a term $t = x_i$, we have $d = 0$ and $n = 1$, and that the claim holds is clear from our choice of $L$. Now, we assume that the claim is true for a depth $d \geq 0$, and we consider a term $t = t_1 \wedge \cdots \wedge t_k$ or a term $t = t_1 \vee \cdots \vee t_k$ having depth $d + 1$. Let $n_i$ denote the number of leaves of $t_i$, and let $d_i$ denote the depth of $t_i$; for each $i$, we have $d_i \leq d$. Also note that since each $t_i$ contains at least one leaf of $t$, we have $k \leq n$. We have

$$\begin{aligned} s(t) &\leq B(s(t_1) + \cdots + s(t_k)) + E \\ &\leq B(u(d_1, n_1) + \cdots + u(d_k, n_k)) + E \\ &\leq B(u(d, n_1) + \cdots + u(d, n_k)) + E \\ &\leq Bnu(d, n) + E \\ &= u(d+1, n). \end{aligned}$$

$\square$

**From Pentagon Entailment to Containment of pp-Formulas.** We now prove Theorem 11.

*Proof (Theorem 11).* Let $\mathbf{A} = (A, \alpha, \beta, \gamma, \mathcal{D})$ be the finite relational structure defined by the finite algebra $\mathbb{A} \in \mathcal{V}(\mathbb{A}_\mathbf{B})$ with universe $A$, the congruences $\alpha, \beta, \gamma \in \text{Con}(\mathbb{A})$, and the finite set of relations $\mathcal{D}$ from Lemma 2. Notice that $\mathbf{A}$ is such that $\text{PPCON}(\mathbf{A}) \in \text{PPCON}(\mathbb{A})$. We claim that $\text{PPCON}(\mathbf{A})$ is coNP-hard, which implies that $\text{PPCON}(\mathbf{B})$ is coNP-hard by Theorem 4. The coNP-hardness of $\text{PPEQ}(\mathbf{B})$ follows from Proposition 1.

Let $\mathcal{P} = \{\mathbf{P}_1, \ldots, \mathbf{P}_{|\mathcal{P}|}\}$ be the finite interesting family of pentagons in Lemma 2. We assume that $P_1$ is the pentagon whose universe is $[\mathbf{a}]/(\beta \wedge \gamma)$, the $\delta$-class that contains the free generator $\mathbf{a}$ of $\mathbb{F}_4$, modulo the congruence $(\beta \wedge \gamma)$.

Let $\phi$ be a sorted pp-formula, and let $\{x_1, \ldots, x_n\}$ and $\{y_1, \ldots, y_m\}$ be the variables of first and second sort in $\phi$, respectively. We let $\{x_1, \ldots, x_{n'}\}$ and $\{y_1, \ldots, y_{m'}\}$ be the free variables of first and second sort in $\phi$, respectively, where $n' \leq n$ and $m' \leq m$. We construct a pp-formula $\phi'$ on $\mathbf{A}$, as follows. For each variable $z$ in $\phi$, we introduce a fresh variable $z'$; $z'$ is existentially quantified in $\phi'$ if and only if $z$ is existentially quantified in $\phi$. If $\phi$ contains the constraint $x_i = x_j$ for some $1 \leq i, j \leq n$, then $\phi'$ contains the conjunct $\beta(x'_i, x'_j)$; if $\phi$ contains the constraint $y_i = y_j$ for some $1 \leq i, j \leq m$, then $\phi'$ contains the conjunct $\gamma(y'_i, y'_j)$; if $\phi$ contains the constraint $R(x_i, y_j, y_k)$ for some $1 \leq i \leq n$ and $1 \leq j, k \leq m$, then $\phi'$ contains the conjunct

$$(\exists w'_1)(\exists w'_2)(\beta(w'_1, x'_i) \wedge \beta(w'_2, x'_i) \wedge \gamma(w'_1, y'_j) \wedge \gamma(w'_2, y'_k) \wedge \alpha(w'_1, w'_2)), \tag{6}$$

where $w'_1$ and $w'_2$ are fresh variables; finally, $\phi'$ contains the conjunct

$$\Delta_{n+m+k}(x'_1, \ldots, x'_n, y'_1, \ldots, y'_m, w'_1, \ldots, w'_k), \tag{7}$$

where $\Delta_{n+m+k}$ is the pp-definition on $\mathbf{A}$ of the relation $D_{n+m+k}$ as in Lemma 2 and $\{w'_1, \ldots, w'_k\}$ is the set of fresh variables introduced in conjuncts of type (6). We arrange $\phi'$ so that the existential quantifiers introduced in conjuncts of type (6) appear at the start of $\phi'$.

The desired reduction is the composition of the reduction $r$ given by Theorem 15 and the mapping $(\phi, \psi) \mapsto (\phi', \psi')$. The construction is feasible in polynomial time by Lemma 2. We show that the reduction is correct.

Let $\mathbf{P}_l$ be a pentagon in $\mathcal{P}$ specified as usual, so that $P_l = B_l \times C_l$ and $P_l \subseteq A$. Let $f$ be a sorted assignment of variables $x_1, \ldots, x_{n'}$ in $B_l$ and $y_1, \ldots, y_{m'}$ in $C_l$, and let $g$ be an assignment of variables $x'_1, \ldots, x'_{n'}$ and $y'_1, \ldots, y'_{m'}$ on $\mathbf{A}$. Say that $f$ and $g$ *match (on $P_l$)* if: $g$ is an assignment in $P_l = B_l \times C_l$; $f(x_i) = b$ if and only if $g(x'_i) = (b, \cdot)$; and, $f(y_i) = c$ if and only if $g(y'_i) = (\cdot, c)$.

*Claim.* If $g$ satisfies $\phi'$ over $\mathbf{A}$ then there is some $\mathbf{P}_l \in \mathcal{P}$ such that $g$ is an assignment over $P_l$ and such that if $f$ is the sorted assignment over $B_l$ and $C_l$ that matches $g$, then $f$ satisfies $\phi$ over $(\mathbf{P}_l)_2$. Conversely, if $\mathbf{P}_l \in \mathcal{P}$ and $f$ is a sorted assignment that satisfies $\phi$ over $(\mathbf{P}_l)_2$ and if $g$ is any assignment over $\mathbf{A}$ that matches $f$, then $g$ satisfies $\phi'$ over $\mathbf{A}$.

*Proof (of Claim).* ($\Rightarrow$) Suppose that $g$ satisfies $\phi'$ over $\mathbf{A}$ and let $g'$ be an extension of $g$ to the quantified variables of $\phi'$ that satisfies each conjunct in $\phi'$. In particular, the sequence of elements given by $g'$ satisfies the conjunct $\Delta_{n+m+k}$ and so belongs to the relation $D_{n+m+k}$. From Lemma 2 it follows that there is some $l$ such that the elements of $g'$ all lie in $P_l$. Let $f$ be the sorted assignment matching $g$ on $P_l = B_l \times C_l$ and let $f'$ be the extension of $f$ to the quantified variables $\{x_i, y_j \mid 1 \leq i \leq n, 1 \leq j \leq m\}$ in $\phi$ defined by $f'(x_i) = b \in B_l$ if and only if $g'(x'_i) = (b, \cdot)$, and $f'(y_j) = c \in C_l$ if and only if $g'(y'_j) = (\cdot, c)$. We prove that $f$ satisfies $\phi$ over $(\mathbf{P}_l)_2$, by checking that $f'$ satisfies each conjunct of $\phi$ over $(\mathbf{P}_l)_2$.

If $g'$ satisfies a conjunct $\beta(x'_i, x'_j)$ in $\phi'$, we want to show that $f'$ satisfies the counterpart $x_i = x_j$ in $\phi$ over $(\mathbf{P}_l)_2$. By Lemma 2, $\beta_l = \beta \cap (P_l)^2$, therefore, $(g'(x'_i), g'(x'_j)) \in \beta_l$, say, $g'(x'_i) = (b, \cdot)$ and $g'(x'_j) = (b, \cdot)$ for some $b \in B_l$. But then $f'(x_i) = f'(x_j) = b$ by the definition of $f'$, and so $f'$ satisfies $x_i = x_j$ in $\phi$ over $(\mathbf{P}_l)_2$. The case of conjuncts of the form $\gamma(y'_i, y'_j)$ in $\phi'$ is similar. If $g'$ satisfies a conjunct of the form in (6) in $\phi'$, we want to show that $f'$ satisfies the counterpart $R(x_i, y_j, y_k)$ in $\phi$ over $(\mathbf{P}_l)_2$. Along the above lines, by direct inspection of (6), the following holds: $g'(x'_i) = (b, \cdot)$, $g'(w'_1) = (b, \cdot)$, and $g'(w'_2) = (b, \cdot)$ for some $b \in B_l$; $g'(y'_j) = (\cdot, c_1)$ and $g'(w'_1) = (\cdot, c_1)$ for some $c_1 \in C_l$; $g'(y'_k) = (\cdot, c_2)$ and $g'(w'_2) = (\cdot, c_2)$ for some $c_2 \in C_l$; and, $((b, c_1), (b, c_2)) \in \alpha_l$. This implies that, if $b$ is in the $r$th block of the partition of $B_l$ defined as in (3), then $(c_1, c_2)$ is in the $r$th congruence induced by $\alpha_l$. But then, by (4), $f'(x_i) = b$, $f'(y_j) = c_1$, and $f'(y_k) = c_2$ imply $(f'(x_i), f'(y_j), f'(y_k)) \in R^{(\mathbf{P}_l)_2}$, that is, $f'$ satisfies $R(x_i, y_j, y_k)$ over $(\mathbf{P}_l)_2$.

($\Leftarrow$) Conversely, suppose that the sorted assignment $f$ satisfies $\phi$ over $(\mathbf{P}_l)_2$. Let $g$ be any assignment on $\mathbf{A}$ matching $f$ on $P_l$, so that by definition, $g(x'_i) = (b, \cdot)$ if and only if $f(x_i) = b \in B_l$, and $g(y'_i) = (\cdot, c)$ if and only if $f(y_i) = c \in C_l$. Let $f'$ be an extension of $f$ to the quantified variables of $\phi$ that satisfies each constraint in $\phi$ over $(\mathbf{P}_l)_2$. Let $g'$ be an extension of $g$ to the quantified variables $x'_{n'+1}, \ldots, x'_n, y'_{m'+1}, \ldots, y'_m$ in $\phi'$ on $P_l = B_l \times C_l$ such that $g'(x'_i) = (b, \cdot)$ if and only if $f'(x_i) = b$ and $g'(y'_i) = (\cdot, c)$ if and only if $f'(y_i) = c$. Note that $g'$ does not assign values to the fresh variables $w'_1$'s and $w'_2$'s arising by (6); Below we check that a suitable extension of $g'$ to such variables satisfies each conjunct of $\phi'$ over $\mathbf{A}$, thus concluding that $g$ satisfies $\phi'$ over $\mathbf{A}$.

Hence, we consider the other conjuncts in $\phi'$, that correspond to constraints in $\phi$ by construction. If $f'$ satisfies the constraint $x_i = x_j$ in $\phi$, that is, $f'(x_i) = f'(x_j) = b$ for some $b \in B_l$, then $g'$ satisfies the counterpart $\beta(x'_i, x'_j)$ in $\phi'$, because $g'(x'_i) = (b, \cdot)$ and $g'(x'_j) = (b, \cdot)$ by definition of $g'$ and $\beta_l = \beta \cap (P_l)^2$ by Lemma 2. The case of constraints of the form $y_i = y_j$ in $\phi$ is similar. If $f'$ satisfies a constraint of the form $R(x_i, y_j, y_k)$ in $\phi$, then by (4), if $f(x_i) = b$ is in the $r$th block in the partition of $B_l$ induced by $\alpha_l$, then $(f'(y_j), f'(y_k)) = (c_1, c_2)$ is in the $r$th congruence induced by $\alpha_l$. Consider the conjunct $\zeta$ of the form in (6) occurring in $\phi'$ by construction. We extend $g'$ to the existentially quantified variables $w'_1$ and $w'_2$ in $\phi'$ by letting $g'(w'_1) = (b, c_1)$ and $g'(w'_2) = (b, c_2)$. By direct inspection, this extension of $g'$ satisfies $\zeta$. Finally, since all of the elements in the assignment $g'$ (extended to the variables $w'_i$) lie in the subset $P_l$, then by Lemma 2, the conjunct (7) is satisfied.

The claim is settled. □

*Claim.* If $(\phi, \psi) \in$ 2-PENTAGON-ENTAILMENT$(\mathcal{P})$ then $(\phi', \psi') \in$ PPCON$(\mathbf{A})$, and if $(\phi, \psi)$ is a no instance of 2-PENTAGON-ENTAILMENT$(\mathcal{P})$ in the range of the reduction given by Theorem 15, then $(\phi', \psi')$ is a no instance of PPCON$(\mathbf{A})$

*Proof (of Claim).* Suppose that $(\phi, \psi) \in$ 2-PENTAGON-ENTAILMENT$(\mathcal{P})$ and let $g$ be any assignment of $\{x'_i, y'_j \mid 1 \leq i \leq n', 1 \leq j \leq m'\}$ that satisfies $\phi'$ over $\mathbf{A}$. By the previous claim, it follows that there is some $\mathbf{P}_l \in \mathcal{P}$ such that $g$ is an assignment over $P_l = B_l \times C_l$ and such that if $f$ is the sorted assignment over $B_l$ and $C_l$ that matches $g$, then $f$ satisfies $\phi$ over $(\mathbf{P}_l)_2$. Since $(\phi, \psi)$ is in 2-PENTAGON-ENTAILMENT$(\mathcal{P})$ then $f$ also satisfies $\psi$ over $(\mathbf{P}_l)_2$ and then by the claim, $g$ must also satisfy $\psi'$ over $\mathbf{A}$ (since $g$ matches $f$). Thus $(\phi', \psi') \in$ PPCON$(\mathbf{A})$.

Now suppose that $(\phi, \psi)$ is a no instance of 2-PENTAGON-ENTAILMENT$(\mathcal{P})$ that lies in the range of the reduction given by Theorem 15. Then, by the Theorem, $\phi \not\models \psi$ is witnessed over any interesting pentagon $\mathbf{P} \in \mathcal{P}$, and in particular by the pentagon $\mathbf{P}_1$ and by some sorted assignment $f$ over $(\mathbf{P}_1)_2$. So, over $(\mathbf{P}_1)_2$, $f$ satisfies $\phi$ and does not satisfy $\psi$. Let $g$ be any assignment over

**A** that matches $f$. By the previous claim, we conclude that $g$ satisfies $\phi'$ over **A**. We argue by contradiction to show that $g$ fails to satisfy $\psi'$. Under the assumption that $g$ satisfies $\psi'$ over **A**, the previous claim establishes that there is some pentagon $\mathbf{P}_l \in \mathcal{P}$ such that $g$ is an assignment over $P_l$ and such that if $f$ is the sorted assignment over $B_l$ and $C_l$ that matches $g$, then $f$ satisfies $\psi$ over $(\mathbf{P}_l)_2$.

We claim that $l = 1$: Since $f$ is a sorted assignment over $(\mathbf{P}_1)_2$ and $g$ matches $f$, then the elements of $g$ lie in $[\mathbf{a}]/(\beta \wedge \gamma)$, the universe of $\mathbf{P}_1$. The pentagon $\mathbf{P}_l$ provided by the previous claim has the property that the elements of $g$ all lie in $\mathbf{P}_l$, and since $P_1$ is the unique pentagon from $\mathcal{P}$ that contains elements from $[\mathbf{a}]/(\beta \wedge \gamma)$, it follows that $l = 1$.

Thus, $f$ satisfies $\psi$ over $(\mathbf{P}_1)_2$, a contradiction, and so we conclude that $g$ does not satisfy $\psi'$ over **A**, establishing that $(\phi', \psi')$ is a no instance of PPCON(**A**). □

Using Theorem 15, we conclude that PPCON(**A**) is coNP-hard, and the proof is complete. □

**Acknowledgements.** Hubie Chen is supported by the Spanish program "Ramon y Cajal" and MICINN grant TIN2010-20967-C04-02. Matt Valeriote acknowledges the support of the Natural Sciences and Engineering Research Council of Canada.